\documentclass[%
 reprint,
superscriptaddress,
 amsmath,amssymb,
 pre,
]{revtex4-2}
\usepackage{centernot}
\usepackage{graphicx}
\usepackage{dcolumn}
\usepackage{bm}
\usepackage{times}
\usepackage{algpseudocodex}
\usepackage{algorithm}

\begin{document}


\title{Effects of clustering heterogeneity on the spectral density of sparse networks}

\author{Tuan Minh Pham}
\affiliation{Niels Bohr Institute, University of Copenhagen,
Blegdamsvej 17, Copenhagen, 2100-DK, Denmark
}

\author{Thomas Peron}
\affiliation{Institute of Mathematics and Computer Science, University of S\~ao Paulo, S\~ao Carlos 13566-590, S\~ao Paulo, Brazil}

\author{Fernando L. Metz}
\affiliation{Physics Institute, Federal University of Rio Grande do Sul, 91501-970 Porto Alegre, Brazil}

\begin{abstract}
  We derive exact equations for the spectral density of sparse networks with an arbitrary distribution of the number
  of single edges and triangles per node. These equations enable a systematic investigation of the effects of
  clustering on the spectral properties of the network adjacency matrix.
  In the case of heterogeneous networks, we
  demonstrate that
  the spectral density becomes more symmetric as the fluctuations in the triangle-degree sequence increase. This phenomenon
  is explained by the small clustering coefficient of networks with a large variance of the triangle-degree distribution.
  In the homogeneous case of regular clustered networks, we find that both perturbative and non-perturbative approximations
  fail to predict the spectral density in the high-connectivity limit. This suggests that traditional large-degree approximations
  may be ineffective in studying the spectral properties of networks with more complex motifs. Our theoretical results
  are fully confirmed by numerical diagonalizations of finite adjacency matrices.
\end{abstract}

\maketitle


\section{Introduction}
The adjacency matrix of a network fully encodes its structure, thus providing  a great amount of information about
dynamical processes on it. This includes the dependence of various critical points on the
adjacency matrix's leading eigenvalue, such as the threshold probability for network percolation \cite{newman2018networks}, the critical infection
rate for epidemic outbreaks \cite{pastor2015epidemic,arruda2018fundamentals}, the onset of synchronization of phase
oscillators \cite{rodrigues2016kuramoto}, and the linear stability of large dynamical systems, such as neural networks \cite{Sompolinsky1988} and ecosystems \cite{may1972will}. 
Structural aspects of networks can also be understood in terms
of the spectral properties of the adjacency matrix. For example,
the eigenvector centrality is the leading eigenvector of the adjacency matrix \cite{newman2018networks}, while the limits in community detection can be expressed in terms of the network spectra \cite{nadakuditi2012graph,peixoto2013eigenvalue}. 
Hence, to understand both the structural and dynamical properties of a network, it is important to determine the spectrum of its adjacency matrix.

The central
question of how the distribution of eigenvalues is affected by a particular network feature, such as the degree heterogeneity, has been studied extensively
using the so-called configuration
model~\cite{farkas2001spectra,chung2003spectra,dorogovtsev2003spectra,rodgers2005eigenvalue,rogers2010spectral,kuhn2011spectra,nadakuditi2013spectra,newman2019pRE,Metz2020}. In this
model, one generates the degrees from a fixed probability distribution and then randomly connects pairs of nodes to match the prescribed degree
sequence. Networks generated through this procedure are locally tree-like, i.e., they contain only long loops of length $O(\ln N)$, where $N$ is the total number of nodes.
Although the configuration model 
allows one to study how degree heterogeneities impact the spectral
properties, the absence of short loops makes this model somewhat unrealistic, as real-world networks often
exhibit considerable levels of \textit{clustering} -- the extent to which 
a triple of nodes form a closed triangle~\cite{newman2018networks,Newman2001}. 

Aiming to overcome this limitation of the configuration model, Newman \cite{Newman2009} and Miller \cite{miller2009percolation} independently introduced a generative network model with tunable clustering.
By generalizing the notion of a degree sequence, each vertex of the Newman-Miller model
is attached to a prescribed number of triangles and single edges (i.e., those edges that do not form triangles). 
This model has been employed to probe the impact of clustering on the spectral properties of networks \cite{Peron_2018,Newman2019, Cantwell2019,Guzman} and  on the  dynamical processes on networks \cite{hackett2011cascades,Volz2011, Huang_2013}.  Approximations  have been derived for the spectral density of networks with a large average number of triangles using
random-matrix techniques  \cite{Peron_2018}, the spectral density of sparse, real-world networks with complex
topologies, including short loops of any length have been determined in \cite{Newman2019,Cantwell2019,Guzman}.

Despite these  advances, the problem of how the spectral density of \emph{sparse} networks (i.e., those with finite average degree) is controlled by the heterogeneity
in the distribution of triangles remains unsolved.
In this paper, we approach this problem by deriving a set of exact self-consistent equations for the full distribution of the diagonal elements of the resolvent, from which  the spectral density can be computed. These equations depend only on
the network's local structure through the joint distribution of the number of single edges and triangles per node, thus generalising the cavity method  developed for Husimi graphs without single edges \cite{bolle2013spectra,Metz2011}. Apart from the spectral density, the resolvent distribution also allows the computations of other spectral observables containing information about the eigenvector statistics \cite{metz2010localization,silva2022analytic,Tapias_2023}. 
 
We illustrate our method for various distributions of the number of single-edges and triangles. While it is well-known that
the skewness of the spectral density depends solely on the number of triangles and edges, and the average degree \cite{vanmieghem2023graph}, we show how, unexpectedly, for a
fixed average number of triangles, fluctuations in the triangle-degree sequence symmetrize the spectral density.  We  explain this phenomenon as a consequence of a decrease in the clustering coefficient with increasing the variance of the triangle distribution. In addition,  for sufficiently
heterogeneous distributions of the number of triangles, the clustering coefficient is no longer a monotonic function of the averaged number of triangles. This implies that
the highest clustering does not occur at the maximum average number of triangles, but at an intermediate value.

In the case of regular
clustered networks with a fixed number of single-edges and triangles, the spectral density converges to the Wigner law 
in the high-connectivity limit, independently of the number of triangles. Moreover, we find that  
perturbative and non-perturbative approximations, both fail to capture the effect of triangles on the spectral density of  networks with large mean degrees. Numerical diagonalization results of finite-size
matrices  confirm our theoretical findings.

This paper is organized as follows. In section II, we describe the random clustered network model as introduced in \cite{Newman2009,miller2009percolation}. In
section \ref{sec:resolvent_equations}, we derive step by step the resolvent equations by generalizing the standard cavity method to networks
with triangles. Section \ref{sec:results} presents analytical and numerical results for the spectral 
density of regular clustered networks and heterogeneous networks. In this section, we also highlight the effects of triangle fluctuations on the spectral density and obtain the
clustering coefficient as a function of the variance of the distribution of triangles. We give our final remarks in section \ref{sec:results}. The paper
contains three appendices:  Appendix \ref{sec:appendix_clustering_coeff} provides the computation of the clustering coefficient,  Appendix~\ref{sec:appendix_correlated} shows the spectral density for networks with correlated degree sequences and
Appendix \ref{sec:appendix_population_dynamics} presents a detailed account of the population dynamics algorithm.


\section{The random clustered network model}

In the Newman-Miller random clustered graph model~\cite{Newman2009,miller2009percolation}, an undirected graph $\mathcal{G}$ of $N$ nodes is generated from
two degree sequences: the sequence $\{s_i\}_{i=1,...,N}$ of single-edge degrees, where $s_i$ specifies the number of edges attached to node $i$ that do
not belong to triangles; and the sequence $\{t_i\}_{i=1,...,N}$, in which $t_i$ is the number of triangles that node $i$ participates in.
Accordingly, the total degree of node $i$ is given by $k_i = s_i + 2 t_i$, as a triangle contributes with two edges to the total degree of a single node.

A single graph instance is generated as follows: first, we draw a joint degree sequence $(s_1,t_1),(s_2,t_2),...,(s_N,t_N)$ from a joint distribution $p_{st}$. We then assign $s_i$ ``stubs'' and $t_i$ ``corners'' of triangles to every node $i$. We create a single network by joining pairs of stubs uniformly at random, and likewise choosing triples of corners also uniformly at random. This process can be implemented using the algorithm in \cite{miller2009percolation} as follows: (i) create two independent lists from the sequence $(s_1,t_1),(s_2,t_2),...,(s_N,t_N)$; (ii) place the label of node $i$ in the first list $s_i$ times and in the second list $t_i$ times; (iii) shuffle the two lists, and then join the pairs of nodes in the positions $2n$ and $2n+1$ of the first list, and the triples in positions $3n$, $3n+1$, and $3n+2$ of the second list. While multiple-edges and self-loops might be created, the probability of their occurrence, as well as discrepancies from the originally imposed degree sequence, become negligible in the limit $N \rightarrow \infty$~\cite{newman2018networks,miller2009percolation}. Other algorithms are able to generate networks with tunable clustering from a fixed total degree sequence $\{k_i\}_{i=1,...,N}$~\cite{heath2011generating}, though correlations between single edges and triangles cannot be set beforehand.

\begin{figure}
\includegraphics[width=1\linewidth]{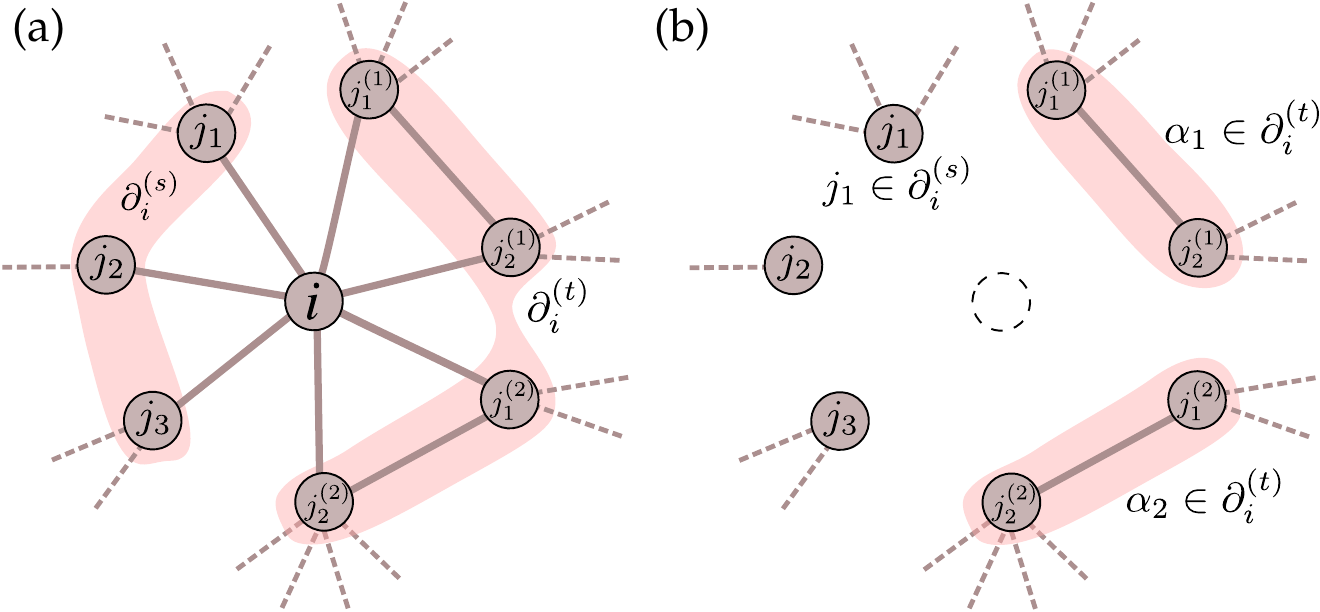}
\caption{Illustration of (a) the original graph $\mathcal{G}$ and (b) a cavity graph $\mathcal{G}^{(i)}$, obtained by removing $i$ together with all of its
  connections from $\mathcal{G}$  in (a). Here the single-edge neighbourhood of $i$ denoted by $\partial_i^{(s)}$ consists of nodes connected to $i$ via
  single-edges: $\partial_i^{(s)} = \left\{ j_1, j_2, j_3\right\}$. The triangle neighbourhood of $i$ denoted by $\partial_i^{(t)}$ is the
  set $\partial_i^{(t)} = \left\{\alpha_1, \alpha_2\right\}$, whose elements $\alpha_1 = \Big\{j_1^{(1)}, j_2^{(1)}\Big\}$ and $\alpha_2 = \Big\{j_1^{(2)}, j_2^{(2)}\Big\}$ are
  the tuples containing  nodes  connected to $i$ via triangle-edges. Upon removal of $i$, the nodes $j_1$, $j_2$, $j_3$ and the tuples $\alpha_1$ and $\alpha_2$, all
  become uncorrelated with each other.}
\label{fig:cavity_method}
\end{figure}

Since there are two different types of edges in the resulting graph $\mathcal{G}$, it is convenient
to split the local neighbourhood of a node into two different sets. Thus, a given node $i$ is adjacent to a subset $\partial_i^{(s)}$ of nodes that define
the single-edge neighbourhood, and to a subset $\partial_i^{(t)}$ of nodes that belong to the triangle-based neighbourhood. More precisely, $\partial_i^{(s)}$ consists of
those nodes that are connected to $i$ by single-edges, while $\partial_i^{(t)}$ is formed by the set of 2-tuples $\alpha = \{j_1^{(\alpha)}, j_2^{(\alpha)}\}$ connected
to $i$ via triangles. Figure \ref{fig:cavity_method}(a) depicts the neighbourhoods $\partial_i^{(s)}$ and  $\partial_i^{(t)}$ around a given node.
The total neighbourhood of $i$ is defined as $\Omega_i \equiv \partial^{(s)}_i \cup \partial^{(t)}_i$. 

The ensemble of graphs is completely specified, in the limit $N \rightarrow \infty$, by the joint distribution
\begin{equation}
    p_{st}= \lim_{N \rightarrow \infty} \frac{1}{N} \sum_{i=1}^N \delta_{s,s_i} \delta_{t,t_i} , 
    \label{eq:pst}
\end{equation}
where $\delta$ represents the Kronecker symbol. The above quantity provides information about fluctuations in the graph topology, since
it gives the fraction of nodes connected to $s$ single-edges and $t$ triangles.
The mean number of single-edges and triangles connected to a node are defined, respectively, by
\begin{equation}
\langle s\rangle = \sum_{s,t=0}^{\infty} s \, p_{st} \quad \text{and} \quad \langle t\rangle = \sum_{s,t=0}^{\infty} t \, p_{st}.
\end{equation}  
The corresponding variances, which quantify the fluctuations in the number of single-edges and triangles per node, are obtained
from $p_{st}$ as follows
\begin{align}
  \sigma_{e}^2 = \sum_{s,t=0}^{\infty} (s-\langle s\rangle)^2 p_{st}, \nonumber \\
  \sigma_{\Delta}^2 = \sum_{s,t=0}^{\infty} (t-\langle t\rangle)^2 p_{st}.
\end{align}  
Due to the local constraint $k_i = s_i + 2 t_i$, the degree
distribution $P_k$ is related to $p_{st}$ as $P_k = \sum_{s,t=0}^{\infty} p_{st} \delta_{k,s+2t}$, while the mean degree $c$ fulfills $c=\langle s\rangle + 2 \langle t\rangle$.
In what follows, we derive a set of self-consistent equations that determine the spectral density of the ensemble of adjacency matrices generated from this model
for an arbitrary distribution $p_{st}$.


\section{The spectral density of the adjacency matrix}
\label{sec:resolvent_equations}

The off-diagonal elements $\{ A_{ij} \}_{i,j=1,\dots,N}$ of the $N \times N$ weighted adjacency matrix $\boldsymbol{A}$ associated to $\mathcal{G}$ store the strength of  symmetric couplings between
pairs $(i,j)$ of adjacent nodes. If $A_{ij}=A_{ji} \neq 0$, there is an undirected edge connecting nodes $i$ and $j$, whereas $A_{ij}=A_{ji}=0$ means  $i$ and $j$ do not
interact. The diagonal elements of $\boldsymbol{A}$ are zero. Since there are two
types of edges in the network, we assume the strength of single-edge interactions is $A_{ij}= J_e \in \mathbb{R}$, while the weight
of the edges belonging to triangles is $A_{ij}= J_\Delta \in \mathbb{R}$. Because $\langle s\rangle$ and $\langle t\rangle$ are finite and independent of $N$, $\boldsymbol{A}$ is
a sparse random matrix with an average number $\langle s\rangle + 2 \langle t\rangle$ of nonzero elements per row or column.

We are interested in the spectral density of the adjacency matrix,
\begin{equation}
  \rho(\lambda) = \lim_{N \rightarrow \infty} \frac{1}{N} \sum_{i=1}^N \delta \left(\lambda - \lambda_i  \right),
  \label{gdqw}
\end{equation}  
where $\lambda_1,\dots,\lambda_N$ denotes the set of real eigenvalues of $\boldsymbol{A}$. Equation (\ref{gdqw}) defines
the most basic spectral observable in random matrix theory, since it provides the average fraction of eigenvalues around $\lambda$.
Below we explain how to derive a set of equations for $\rho(\lambda)$.


\subsection{Mapping the problem into a statistical physics calculation}

By introducing the $N \times N$ resolvent matrix $\boldsymbol{G}(z)\equiv (z \boldsymbol{I} -\boldsymbol{A})^{-1}$, with $\boldsymbol{I}$ denoting
the identity matrix and $z = \lambda - i \varepsilon  \in \mathbb{C}$ ($\varepsilon > 0$), the
spectral density $\rho(\lambda)$ can be expressed in terms of $\boldsymbol{G}(z)$ via the inverse Stieltjes transform \cite{livan2018introduction}
\begin{equation}
    \rho(\lambda) = \lim_{\varepsilon \rightarrow 0^+}\lim_{ N \rightarrow \infty}\, \frac{1}{\pi N}\, {\rm Im}\,  {\rm Tr}\, \boldsymbol{G}(z).
\label{eq:spectral_density}
\end{equation}
Clearly, by computing the diagonal elements $\{ G_{ii}(z) \}_{i=1,\dots,N}$ of the resolvent matrix, we are able to reconstruct $\rho(\lambda)$ from the above equation.

To compute the diagonal elements $G_{11}(z),\dots,G_{NN}(z)$, we use the cavity approach as developed in random
matrix theory \cite{Rogers2008,metz2010localization}. This method, upon introducing
site-dependent real variables $\{ x_i \}_{i=1,\dots,N}$ and an associated Hamiltonian  
\begin{equation}
H_z(\boldsymbol{x}) = \frac{i}{2}\, \sum_{i,j = 1}^N x_i (z \boldsymbol{I} -\boldsymbol{A})_{ij}x_j\,,
\label{eq:Hamiltonian}
\end{equation}
recasts the problem of determining $\rho(\lambda)$ into a statistical mechanics problem of $N$ interacting variables placed at the nodes of $\mathcal{G}$. In fact, by
representing $G_{ii}(z) = (z \boldsymbol{I} -\boldsymbol{A})_{ii}^{-1}$ as a Gaussian integral over $\{ x_i \}_{i=1,\dots,N}$, we rewrite Eq. (\ref{eq:spectral_density}) as
\begin{equation}
  \rho(\lambda) = \lim_{\varepsilon \rightarrow 0^+}\lim_{ N \rightarrow \infty}\, \frac{1}{\pi N}\,\sum_{k=1}^N
      {\rm Re} \left[ \int\limits_{-\infty}^{\infty} \left( \prod_{i=1}^N d x_i  \right) x_k^2 \, \mathcal{P}_z(\boldsymbol{x}) \right],
      \label{furwa}
\end{equation}
where we introduced the auxiliary joint distribution
\begin{equation}
\mathcal{P}_z(\boldsymbol{x}) = \frac{e^{-H_z(\boldsymbol{x})}}{ \int\limits_{-\infty}^{\infty}  \left( \prod_{i=1}^N d x_i  \right)   \, e^{-H_z(\boldsymbol{x})}}
\end{equation}  
of $\boldsymbol{x}= (x_1,\dots,x_N)$. We conclude that the second moments of the
local marginals $\mathcal{P}_z(x_k) = \int_{-\infty}^{\infty} \left( \prod_{i=1 (\neq k)}^N d x_i \right) \mathcal{P}_z(\boldsymbol{x})$ ($k=1,\dots,N$)  determine the spectral density by means
of Eq. (\ref{furwa}).

\subsection{Determining the local marginals with the cavity method}

Here we explain how to derive a set of self-consistent equations for the local marginals $\mathcal{P}_z(x_k)$ ($k=1,\dots,N$). Let $\boldsymbol{x}_{\mathcal{N}}$ represent
the variables $\{ x_i \}_{i \in \mathcal{N}}$ on a particular subset $\mathcal{N}$ of nodes. We also define $d \boldsymbol{x}_{\mathcal{N}} = \prod_{i \in \mathcal{N}} d x_i$
as the integration measure over $\boldsymbol{x}_{\mathcal{N}}$.
In accordance to the partition of $\Omega_k$  into $\partial_k^{(s)}$ and  $\partial_k^{(t)}$, the graph Hamiltonian $H_z(\boldsymbol{x})$ can be decomposed as
\begin{equation}\label{eq:hamiltonian_Hz}
    H_z(\boldsymbol{x}) =  \frac{i}{2}  \sum_{k = 1}^N \left[z \, x_k^2  -  x_k h_k\left( \boldsymbol{x}_{\Omega_k} \right)  \right],  
\end{equation}
where the local field $h_k\left( \boldsymbol{x}_{\Omega_k} \right)$ at site $k$ reads
\begin{equation}
  h_k\left( \boldsymbol{x}_{\Omega_k} \right)  = J_e \sum_{j \in \partial_k^{(s)}} x_j + J_\Delta \sum_{\alpha \in \partial_k^{(t)}} \boldsymbol{x}_{\alpha}^{T}  \boldsymbol{u},
  \label{eq:local_field} 
\end{equation}
with $\boldsymbol{u}$ representing the two-dimensional uniform vector $\boldsymbol{u}^{T} = (1 \,\,\,\, 1)$. The vector $\boldsymbol{x}_\alpha^{T} = (x_{j_1}^{(\alpha)}, \,\, x_{j_2}^{(\alpha)})$ refers to the
pair of variables corresponding to a tuple $\alpha \in \partial_k^{(t)}$.
Note that $h_k\left( \boldsymbol{x}_{\Omega_k} \right)$ is independent of  $x_k$. Notice also that the graph Hamiltonian in Eq.~\eqref{eq:hamiltonian_Hz} can be rewritten as
\begin{equation}
    H_z(\boldsymbol{x}) = \frac{i}{2} z \, x_k^2 - i x_k h_k\left( \boldsymbol{x}_{\Omega_k} \right)  + H^{(k)}_z(\boldsymbol{x}), 
\end{equation}
where the first term on the right-hand side is the contribution of node $k$ to $H_z(\boldsymbol{x})$, while $H^{(k)}_z(\boldsymbol{x})$ 
is the Hamiltonian on the cavity graph $\mathcal{G}^{(k)}$, which is the graph obtained from $\mathcal{G}$ by removing node $k$ and all its edges [see  Figure 1(b)].
This allows us to express the marginal distribution $\mathcal{P}_z(x_k)$ as follows
\begin{equation} 
\mathcal{P}_z(x_k) \propto e^{  - \frac{i z }{2} x_k^2 }  \int d\boldsymbol{x}_{\Omega_k}
\mathcal{P}^{(k)} \big(\boldsymbol{x}_{\Omega_k} \big)   \exp\left[ ix_k h_k\left( \boldsymbol{x}_{\Omega_k} \right)  \right] , 
\label{eqn:cavity1}
\end{equation}
where $\mathcal{P}^{(k)} \big(\boldsymbol{x}_{\Omega_k}\big)$ is the joint distribution of $\boldsymbol{x}_{\Omega_k}$ on $\mathcal{G}^{(k)}$.
We have ignored an unimportant normalization factor when writing  Eq. (\ref{eqn:cavity1}).

The fundamental property that allows us to compute the spectral density using the cavity method is the local tree-like topology
of the Newman-Miller model of networks \cite{miller2009percolation,Newman2009}. 
This property implies that correlations among the neighbours of a given node $i$ are mediated mainly through $i$ [see Figure 1(a)]. After removing
the central node $i$, the neighbourhood $\Omega_i$ splits into different subsets of
nodes that are essentially disconnected from each other, as illustrated in Figure 1(b). For large $N$, the typical length of a
path that connects two nodes belonging to distinct subsets in $\Omega_i$ is of $\mathcal{O}(\ln N)$ \cite{Bordenave2010,Metz2011}. Thus, the nodes $j\in \partial_i^{(s)}$ and
the tuples $\alpha \in \partial_i^{(t)}$ become mutually independent on the cavity graph $\mathcal{G}^{(k)}$, yet
nodes belonging to the same tuple remain correlated. 

Therefore, since the distribution $\mathcal{P}^{(k)} \left( \boldsymbol{x}_{\Omega_k} \right)$ is defined on the cavity graph $\mathcal{G}^{(k)}$, the local tree-like
structure of $\mathcal{G}$ leads to the factorization property  \cite{Kikuchi1951,yedidia2001bethe}
\begin{equation}
  \mathcal{P}_z^{(k)} \left( \boldsymbol{x}_{\Omega_k} \right) =  \prod_{j \in \partial_k^{(s)}} \mathcal{P}_z^{(k)}(x_j) \prod_{ \alpha \in \partial_k^{(t)}} \mathcal{P}_z^{(k)}(\boldsymbol{x}_\alpha).
\label{factorisation}
\end{equation}
Equation (\ref{factorisation}) lies at the core of the cavity
method since it enables writing $\mathcal{P}_z(x_k)$ in terms of local distributions defined
on the cavity graph. Indeed, by substituting Eq.~(\ref{factorisation}) in Eq.~(\ref{eqn:cavity1}) and using Eq.~(\ref{eq:local_field}), we find
\begin{align} 
  &\mathcal{P}_z(x_k) \propto e^{  - \frac{i z }{2} x_k^2 }  \prod_{j \in \partial_k^{(s)}}  \int_{-\infty}^{\infty} d x_j \mathcal{P}_z^{(k)}(x_j) \exp{\left( i J_e x_k x_j \right)} \nonumber \\
  &\times
 \prod_{ \alpha \in \partial_k^{(t)}}  \int_{-\infty}^{\infty} d \boldsymbol{x}_\alpha \mathcal{P}_z^{(k)}(\boldsymbol{x}_\alpha)  \exp{\left( i J_\Delta x_k \, \boldsymbol{x}_{\alpha}^{T}  \boldsymbol{u} \right)} .
\label{eqn:cavity2}
\end{align}
Thus, the local marginals $\mathcal{P}_z^{(k)}(x_j)$ and $\mathcal{P}_z^{(k)}(\boldsymbol{x}_\alpha)$ on the cavity graph determine $\mathcal{P}_z (x_k)$ through
the above equation. The rest of the calculation consists of deriving self-consistent equations for $\mathcal{P}_z^{(k)}(x_j)$ and $\mathcal{P}_z^{(k)}(\boldsymbol{x}_\alpha)$.

The marginal distribution $\mathcal{P}_z^{(k)}(x_j)$ on $\mathcal{G}^{(k)}$, with $k\in \partial_j^{(s)}$, is computed by departing from an expression analogous to Eq.~(\ref{eqn:cavity1}), namely
\begin{equation} 
\mathcal{P}_z^{(k)} (x_j) \propto e^{- \frac{i z}{2} x_j^2  }   \int d\boldsymbol{x}_{\Omega_j \backslash k  }
\mathcal{P}_z^{(j,k)} \left( \boldsymbol{x}_{ \Omega_j \backslash k   } \right)
e^{ix_j h_j \left( \boldsymbol{x}_{\Omega_j \backslash k} \right) } , 
\label{eqn:cavity2a}
\end{equation}
in which $\Omega_j \backslash k \equiv  \big(\partial_j^{(s)} \backslash k \big)\cup \partial^{(t)}_j$ represents the neighbourhood of node $j$ except
for $k\in \partial_j^{(s)}$. The distribution $\mathcal{P}_z^{(j,k)} \left( \boldsymbol{x}_{   \Omega_j \backslash k    } \right)$ is defined on a graph $\mathcal{G}^{(j,k)}$ where
both nodes, $j$ and $k\in \partial_j^{(s)}$, and all their incident edges are absent. By invoking again the local tree-like topology of the network, one concludes that
$\mathcal{P}_z^{(j,k)} \big(\boldsymbol{x}_{ \Omega_j \backslash k   } \big)$ factorizes as follows
\begin{equation}
  \mathcal{P}_z^{(j,k)} \left( \boldsymbol{x}_{   \Omega_j \backslash k    } \right) =
  \prod_{\ell \in \partial_j^{(s)} \backslash k  } \mathcal{P}_z^{(j)}(x_\ell) \prod_{ \alpha \in \partial_j^{(t)}} \mathcal{P}_z^{(j)}(\boldsymbol{x}_\alpha).
\label{factorisation1}
\end{equation}
Note that the local marginals on the right-hand side of the above equation are defined on $\mathcal{G}^{(j)}$ instead of $\mathcal{G}^{(j,k)}$. Indeed, given that the variables
on the neighbourhood $\partial_j^{(s)}  \backslash k $ are mainly correlated through $j$, they become independent by simply deleting node $j$ and all its edges. In other
words, the removal of the extra
node $k \in \partial_j^{(s)}$ is irrelevant for the factorization of $\boldsymbol{x}_{\partial_j^{(s)}  \backslash k  }$.
This is a crucial step to close the cavity equations and find the expression
\begin{align} 
  &\mathcal{P}_z^{(k)} (x_j) \propto e^{  - \frac{i z }{2} x_j^2 }  \prod_{\ell \in \partial_j^{(s)} \backslash k}  \int_{-\infty}^{\infty} d x_\ell \mathcal{P}_z^{(j)}(x_\ell) \exp{\left( i J_e x_j x_\ell \right)} \nonumber \\
  &\times
 \prod_{ \alpha \in \partial_j^{(t)}}  \int_{-\infty}^{\infty} d \boldsymbol{x}_\alpha \mathcal{P}_z^{(j)}(\boldsymbol{x}_\alpha)  \exp{\left( i J_\Delta x_j \, \boldsymbol{x}_{\alpha}^{T}  \boldsymbol{u} \right)},
\label{hfgsa}
\end{align}
which is obtained by inserting Eq. (\ref{factorisation1}) into Eq. (\ref{eqn:cavity2a}).

At last, we need to derive an equation
for the local marginal $\mathcal{P}_z^{(k)}(\boldsymbol{x}_\alpha)$, with $\alpha = \{j^{(\alpha)}_1, j^{(\alpha)}_2 \} \in \partial_k^{(t)}$. This is achieved by considering a
graph $\mathcal{G}^{(k,\alpha)}$ obtained from $\mathcal{G}^{(k)}$ by
removing the tuple $\alpha = \{j_1, j_2 \}$, together with all edges incident to $j_1$ and $j_2$ (for simplicity, we have dropped the dependency
of $j_{1}^{(\alpha)}$ and $j_{2}^{(\alpha)}$ with respect to $\alpha$). The nodes $\{ k,j_1,j_2 \}$
form a triangle in the original graph $\mathcal{G}$.
Thus, by repeating the same reasoning as developed above and using the local tree-like topology of the
neighbourhoods $\Omega_{j_1}$ and $\Omega_{j_2}$, we obtain
\begin{align}
  \nonumber
  &\mathcal{P}_z^{(k)}  (\boldsymbol{x}_\alpha)  \propto  \exp\left[ -\frac{iz}{2}\, \big(x^2_{j_1} + x^2_{j_2}\big)  + i  J_\Delta x_{j_1} x_{j_2}  \right] \\
  &\times \prod_{l\in \partial_{j_1}^{(s)} } \int_{-\infty}^{\infty} dx_l \, \mathcal{P}_z^{(j_1)}(x_l)  \exp \left( i J_e  x_{j_1} x_l \right) \nonumber \\
  &\times  \prod_{n \in \partial_{j_2}^{(s)}  } \int_{-\infty}^{\infty} dx_n \, \mathcal{P}_z^{(j_2)}(x_n) \exp \left( i J_e x_{j_2}  x_{n}  \right) \nonumber \\
  &\times  \prod_{\beta\in \partial_{j_1}^{(t)} \backslash \alpha_1} \int_{-\infty}^{\infty}   d\boldsymbol{x}_\beta  \mathcal{P}_z^{(j_1)}(\boldsymbol{x}_\beta)
  \exp \left( i J_\Delta  x_{j_1}  \boldsymbol{x}^{T}_{\beta}  \boldsymbol{u} \right) \nonumber \\
  &\times  \prod_{\gamma \in \partial_{j_2}^{(t)} \backslash \alpha_2}   \int_{-\infty}^{\infty} d\boldsymbol{x}_{\gamma}
  \mathcal{P}_z^{(j_2)} (\boldsymbol{x}_\gamma)  \exp \left( i J_\Delta  x_{j_2}  \boldsymbol{x}^{T}_{\gamma}  \boldsymbol{u}  \right),
\label{eq:newman_resolven3}
\end{align}
with $\alpha_1 = \{j_2, k \}$ and $\alpha_2 = \{j_1, k \}$.
Equations (\ref{hfgsa}) and (\ref{eq:newman_resolven3}) form a closed system of equations for the local
marginals $\mathcal{P}_z^{(k)} (x_j)$ and $\mathcal{P}_z^{(k)} (\boldsymbol{x}_\alpha)$ on the cavity graph $\mathcal{G}^{(k)}$. 

Equations (\ref{hfgsa}) and (\ref{eq:newman_resolven3}) are also known as belief-propagation or message-passing equations \cite{MezardBook}, since
$\mathcal{P}_z^{(k)} (x_j)$ and $\mathcal{P}_z^{(k)} (\boldsymbol{x}_\alpha)$ can be thought of as representing messages that propagate locally among different regions of the graph. In this
context, $\mathcal{P}_z^{(k)} (x_j)$ is a message that propagates from node $j$ to node $k$ through a single-edge, while $\mathcal{P}_z^{(k)} (\boldsymbol{x}_\alpha)$
is a message that propagates from tuple $\alpha$ to node $k$ by following a pair of edges. The interpretation of Eqs. (\ref{hfgsa}) and (\ref{eq:newman_resolven3})
as a message-passing algorithm provides a recursive way to efficiently calculate the local marginals on a graph and determine the spectral density for finite $N \gg 1$.

From Eqs. (\ref{hfgsa}) and (\ref{eq:newman_resolven3}), we can finally derive a set of equations for the diagonal elements $\{ G_{kk}(z) \}_{k=1,\dots,N}$ of the resolvent.
By inspecting the form of Eqs. (\ref{hfgsa}) and (\ref{eq:newman_resolven3}), one infers that an appropriate Gaussian {\it ansatz} for the local marginals solves the
cavity equations. We thus assume that the local marginals are given by \cite{Rogers2008,metz2010localization}
\begin{align}
&\mathcal{P}_z(x_k) = \sqrt{\frac{i}{2 \pi G_{kk}(z)}}\, \exp{\left( - \frac{i  x_k^2}{2G_{kk}(z)} \right) }, \\
&\mathcal{P}_z^{(k)}(x_j) = \sqrt{\frac{i}{2 \pi G_{jj}^{(k)}(z)}}\, \exp{\left(- \frac{i x_j^2}{2G_{jj}^{(k)}(z)} \right) },  \\
&\mathcal{P}_z^{(k)}(\boldsymbol{x}_\alpha) = \sqrt{\frac{i}{2 \pi \, {\rm det} \big[ \mathbb{G}_{\alpha}^{(k)}(z) \big]  }}\,
e^{- \frac{i }{2} \boldsymbol{x}_{\alpha}^{T} . \left[ \mathbb{G}_{\alpha}^{(k)}(z) \right]^{-1} \boldsymbol{x}_\alpha },
\label{eqn:cavity4}
\end{align}
where $\{ G_{jj}(z) \}$ and $\{ G_{jj}^{(k)}(z) \}$ are the diagonal elements of the resolvent defined on $\mathcal{G}$ and $\mathcal{G}^{(k)}$, respectively. The $2 \times 2$
complex matrix $\mathbb{G}_{\alpha}^{(k)}(z)$ is the resolvent associated to a tuple $\alpha \in \partial_{k}^{(t)}$ of the cavity graph $\mathcal{G}^{(k)}$.
Substituting the above assumptions in Eqs.~(\ref{eqn:cavity2}),~(\ref{hfgsa}) and~(\ref{eq:newman_resolven3}), and computing the
Gaussian integrals, we find an expression for $G_{kk}(z)$,
\begin{equation}
  \left[G_{kk}(z)\right]^{-1} = z - J_e^2 \sum_{j \in \partial_k^{(s)}}  G^{(k)}_{jj} - J_\Delta^2 \sum_{\alpha \in \partial_k^{(t)}} \boldsymbol{u}^{T}  \mathbb{G}^{(k)}_{\alpha} \boldsymbol{u},
  \label{juuw}
\end{equation}  
where the cavity resolvents, $\{ G_{jj}^{(k)} \}$ and $\{ \mathbb{G}^{(k)}_{\alpha}   \}$, solve the self-consistent equations
\begin{align}
&\big[ G_{kk}^{(l)} \big]^{-1} = z - J_{e}^2 \sum_{j \in \partial_k^{(s)} \backslash l}  G^{(k)}_{jj} - J_\Delta^2
  \sum_{\alpha \in \partial_k^{(t)}}  \boldsymbol{u}^{T}   \mathbb{G}^{(k)}_{\alpha}  \boldsymbol{u}, \label{jj1} \\
  &\big[   \mathbb{G}^{(k)}_{\alpha}  \big]^{-1}=  z \mathbb{I}  -  \mathbb{A} - \mathbb{D}^{(k)}_{\alpha},
    \label{eq:resolvent1}
\end{align}
with $\mathbb{I}$ denoting the $2 \times 2$ identity matrix.
The elements of the two-dimensional matrices $\mathbb{A}$ and $\mathbb{D}^{(k)}_{\alpha}$ are given by
\begin{align}
  & (\mathbb{A})_{ij} = J_\Delta (1-\delta_{ij}), \nonumber  \\
  &  \big( \mathbb{D}^{(k)}_{\alpha} \big)_{ij} = \delta_{ij} \Bigg( \displaystyle  J_\Delta^2 \sum_{\beta \in \partial_{j_i}^{(t)}\backslash \alpha_i }
    \boldsymbol{u}^{T}  \mathbb{G}_\beta^{(j_i)}  \boldsymbol{u}  + J_{e}^2 \sum_{l \in \partial_{j_i}^{(s)}  }  G^{(j_i)}_{ll} \Bigg). \nonumber
 \label{eq:A_D11_D22}
\end{align}
The resolvent Eqs. (\ref{juuw}-\ref{eq:resolvent1})  provide an approximation
for the spectral density of a single network realization drawn
from the Newman-Miller model with a large number $N$ of nodes. Once we obtain a solution of Eqs. (\ref{jj1}) and (\ref{eq:resolvent1}), the diagonal elements
of the resolvent follow from Eq. (\ref{juuw}).


\subsection{The resolvent distributional equations}

Equations (\ref{juuw}-\ref{eq:resolvent1}) become asymptotically exact as $N \rightarrow \infty$, thanks to
the local tree-like topology of the network. In the limit $N \rightarrow \infty$, instead of solving Eqs.~(\ref{juuw}-\ref{eq:resolvent1}) on a single graph instance, it
is sensible to work with an ensemble
of random graphs and compute the probability density $W_z(g)$ of the diagonal elements of the resolvent. According
to Eq.~(\ref{eq:spectral_density}), the spectral density can be written in terms of $W_z(g)$ as
\begin{equation}
  \rho(\lambda) = \frac{1}{\pi} \lim_{\epsilon \rightarrow 0^+} \int_{\mathbb{H}^+} d g W_z(g) {\rm Im} g,
  \label{gspoq}
\end{equation}  
with $dg = d {\rm Re} g \, d {\rm Im} g$. The symbol $\mathbb{H}^+$ denotes the upper half complex plane. The
probability density $W_z(g)$ that a  given  diagonal element $G_{kk}(z)$ is equal to $g$ reads
\begin{equation}
W_z(g) = \lim_{N \rightarrow\infty} \frac{1}{N}\,  \sum_{k =1}^N \delta\big[g - G_{kk}(z)\big],  
\label{eq:distributional1}
\end{equation}
where the Dirac-$\delta$ distribution here refers to the complex plane. 
Likewise, we introduce the probability densities $W_{e}(g)$ and $W_{\triangle}(\mathbb{G})$ of the
resolvents $G_{jj}^{(k)}$ and $\mathbb{G}_{\alpha}^{(k)}$, respectively. Such empirical distributions are formally defined as
\begin{equation}
W_{e}(g) = \lim_{N \rightarrow\infty} \frac {\sum_{j =1}^N \sum_{i\in \partial_j^{(s)}} \delta \big( g - G_{jj}^{(i)}\big)}{\sum_{i=1}^N s_i  } 
\label{eq:distributional2}
\end{equation}
and
\begin{equation}
  W_{\triangle}(\mathbb{G}) =\lim_{N \rightarrow\infty} \,\frac{\sum_{i=1}^N \sum_{\alpha\in\partial_i^{(t)}}
    \delta \big( \mathbb{G} -\mathbb{G}_{\alpha}^{(i)}  \big) }{\sum_{i=1}^N t_i}.
\label{eq:distributional3}
\end{equation}
The normalization factors in the above equations are related to the total number $N_e$ of edges and to the total
number $N_{\Delta}$ of triangles as $ N_e = \frac{1}{2} \sum_{i=1}^N s_i$ and $N_{\Delta} = \frac{1}{3} \sum_{i=1}^N t_i $.
Note that $W_{e}(g)$ and $W_{\triangle}(\mathbb{G})$  depend on $z \in \mathbb{C}$, but we have omitted such dependency for the sake
of simplifying the notation.

Our aim is to derive a set of self-consistent equations that determine $W_z(g)$ and, consequently, the spectral density $\rho(\lambda)$.
There are two ways of obtaining such equations, either by inserting Eqs. (\ref{juuw}-\ref{eq:resolvent1}) into the
Dirac-$\delta$ distributions of the above definitions and then computing the ensemble averages, or by simply inspecting what network features control
the statistics of the resolvents and then requiring
self-consistency, as dictated by Eqs. (\ref{juuw}-\ref{eq:resolvent1}). Here we follow the second approach, which is rooted in the fact that the two sides
of each one of Eqs.~(\ref{juuw}-\ref{eq:resolvent1}) must be equal in a distributional sense.

The statistics of  $G_{jj}$, $G_{jj}^{(i)}$ and $\mathbb{G}_{\alpha}^{(i)}$ is determined by the probability $p_{st}$ that a node has $s$ single edges and $t$ tuples.
The resolvent $G_{jj}$ is a single-node quantity that depends on $p_{st}$.
The distribution of the single-edge resolvent $G_{jj}^{(i)}$ is determined by the probability $s p_{st}/\langle s\rangle$, which is the distribution of the number of edges and triangles attached to a node reached by a single edge. The quantity $p_{st} t/ \langle t \rangle$ is the analogous distribution for a node reached by a triangle~\cite{Newman2009}. Finally, $\mathbb{G}_{\alpha}^{(i)}$ is a triangle-edge  quantity which depends on the probability $t_1t_2p_{s_1t_1} p_{s_2 t_2}/\langle t\rangle^2$
to select an edge from a triangle such that one of its nodes is adjacent to $(s_1,t_1)$ and the other to $(s_2,t_2)$. 
By requiring that the two sides of each one of Eqs. (\ref{juuw}-\ref{eq:resolvent1}) are equal
in distribution, we find an expression for the probability density of the resolvent
\onecolumngrid
\begin{equation}
 W_{z}(g) =\sum_{s,t=0}^{\infty} p_{st} \int \left( \prod_{\ell=1}^s dg_{\ell} W_e(g_{\ell}) \right) \int \left(\prod_{n=1}^t d \mathbb{G}_n W_{\triangle}(\mathbb{G}_n) \right)
 \delta \Bigg[ g - \left( z - J_e^2\sum_{\ell=1}^s  g_{\ell} - J_\Delta^2\sum_{n=1}^t \boldsymbol{u}^T \mathbb{G}_n \boldsymbol{u}  \right)^{-1}  \Bigg]
 \label{green}
\end{equation}  
where $W_e(g)$ and $W_{\triangle}(\mathbb{G})$ follow from the solutions of the self-consistent equations

\begin{align}
 &W_e(g) =\sum_{s,t=0}^{\infty} \frac{s p_{st}}{\langle s\rangle} \int\left[ \prod_{\ell=1}^{s-1} dg_{\ell} W_e(g_{\ell}) \right] \left[\prod_{n=1}^t  d \mathbb{G}_n W_{\triangle}(\mathbb{G}_n) \right]
  \delta \Bigg[ g - \left( z - J_e^2\sum_{\ell=1}^{s-1}  g_{\ell} - J_\Delta^2 \sum_{n=1}^t \boldsymbol{u}^T \mathbb{G}_n \boldsymbol{u}  \right)^{-1} \Bigg], \label{distributional1}
\\
&W_{\triangle}(\mathbb{G}) =\sum_{s_1,t_1, s_2, t_2=0}^{\infty} \frac{t_1 \, t_2 \, p_{s_1t_1} \, p_{s_2 t_2}}{\langle t \rangle^2} \int
\left[\prod_{\ell=1}^{s_1} dg_{1,\ell} W_e(g_{1,\ell}) \right] \int \left[\prod_{\ell=1}^{s_2} dg_{2,\ell} W_e(g_{2,\ell}) \right]   \nonumber \\
&   \times  \int \left[\,\prod_{n=1}^{t_1-1}  d \mathbb{G}_{1,n} W_{\triangle}(\mathbb{G}_{1,n}) \right]
\int \left[\,\prod_{n=1}^{t_2-1} d\mathbb{G}_{2,n} W_{\triangle}(\mathbb{G}_{2,n}) \right]
\delta \left[ \mathbb{G}  - \left( z \mathbb{I} - \mathbb{A} - \mathbb{D}  \right)^{-1}   \right],
 \label{distributional}
\end{align}
with the two-dimensional matrix $\mathbb{D}$ defined as 
\twocolumngrid
\begin{equation}
  (\mathbb{D})_{ij} = \delta_{ij} \left( J^2_e\sum_{\ell=1}^{s_i}  g_{i,\ell} + J^2_\Delta\sum_{n=1}^{t_i-1} \boldsymbol{u}^T \mathbb{G}_{i,n} \boldsymbol{u} \right). 
\end{equation} 
For the sake of simplicity, we have omitted the dependency of $\mathbb{D}$ in Eq. (\ref{distributional}) with respect to the integration variables.

Equations  (\ref{green}-\ref{distributional}) make up the main analytic result of this work, since the solutions for $W_z(g)$ determine the spectral
density of the Newman-Miller model in the limit $N \rightarrow \infty$ (see Eq.~(\ref{gspoq})).
In the next section, we use the population dynamics algorithm \cite{Rogers2008,metz2010localization} to sample the different terms appearing on the right-hand side of Eqs. (\ref{green}-\ref{distributional}) and iteratively solve these equations.
A detailed account of the population dynamics algorithm is presented in appendix \ref{sec:appendix_population_dynamics}.

  
\section{Results}
\label{sec:results}

In the previous section, we obtained the resolvent distributional Eqs.~(\ref{green}-\ref{distributional}) for the spectral density $\rho(\lambda)$ of clustered networks with an arbitrary
distribution $p_{st}$ of single-edges and triangles. Here we use those equations to study $\rho(\lambda)$
in two particular cases: random regular clustered graphs and networks with heterogeneous distributions of triangles.
\begin{figure}
  \centering  \includegraphics[scale=0.6]{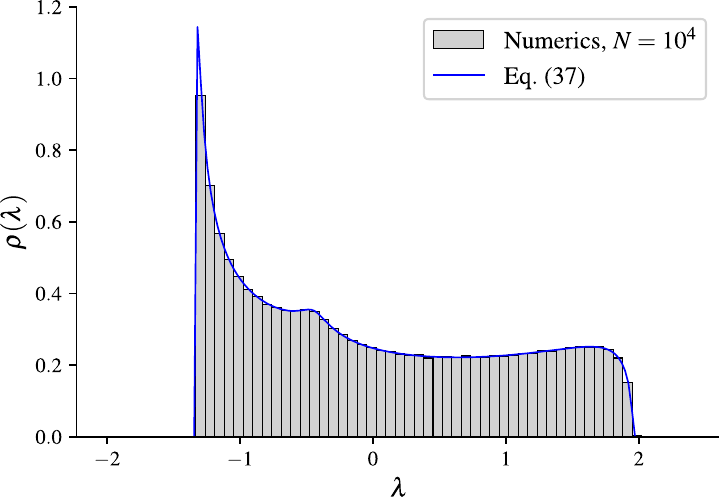}
  \caption{The spectral density $\rho(\lambda)$ of a random regular clustered network with $\langle s\rangle=1$ single-edges and $\langle t\rangle = 2$ triangles. The weights
    of the two different types of edges are $J_e=J_\Delta=1$. The solid line depicts the theoretical result obtained by solving the quartic Eq.~(\ref{eq:quartic}) for $G_\Delta$. The histogram is
    constructed by numerically diagonalizing the adjacency matrix of a single network realization with $N=10^4$ nodes.}
  \label{fig:regular_net}
\end{figure} 
\begin{figure}
  \centering
  \includegraphics[scale=0.65]{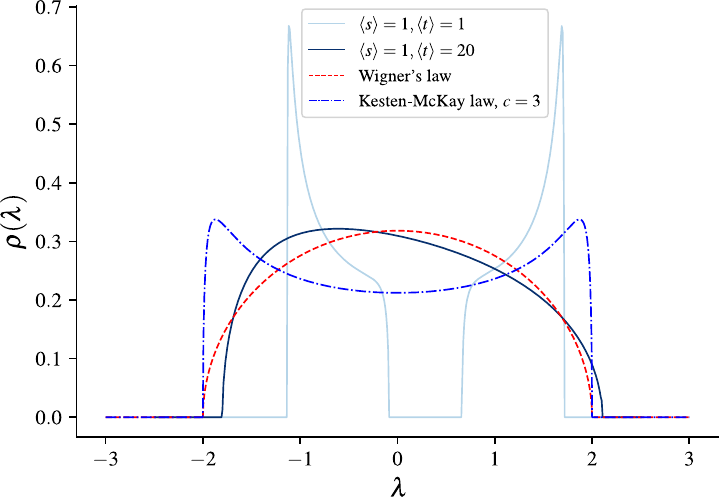}
  \caption{The spectral density $\rho(\lambda)$ of a random regular clustered network with $\langle s\rangle$ single-edges and $\langle t\rangle$ triangles. The weights
    of the two different types of edges are $J_e=J_\Delta=1$.
    The solid lines show the theoretical results obtained by numerically solving Eqs. (\ref{eq:resolvent3}-\ref{eq:resolvent5}) for different
    values of $\langle s\rangle$ and $\langle t\rangle$.
    The dashed line corresponds to
    the Wigner semicircle law of random matrix theory, which is recovered when the network mean degree is infinitely large. The dash-dotted
    line is the spectral density of a random regular network in which every node is connected to $\langle s\rangle=3$ single-edges (Kesten-McKay law).}
  \label{fig:regular_net1}
\end{figure}


\subsection{Random regular clustered networks}\label{sec:regular_networks}

If every node has exactly $\langle s\rangle$ single-edges and $\langle t\rangle$ triangle-edges, then $p_{s t} = \delta_{s,\langle s\rangle} \delta_{t, \langle t\rangle} $. In this
case, the network becomes homogeneous and the solutions of Eqs. (\ref{green}-\ref{distributional}) read
\begin{align}
  &W_z(g) = \delta(g-G), \nonumber \\
  &W_{e}(g) = \delta(g - G_e), \nonumber \\
  &W_{\triangle}(\mathbb{G}) = \delta(\mathbb{G} - \mathbb{G}_\Delta).
\end{align}
By inserting the above expressions into Eqs. (\ref{green}-\ref{distributional}), we conclude that $G$, $G_e$ and $\mathbb{G}_t$ fulfill
the algebraic equations
\begin{eqnarray}
     G^{-1} &=& z - \langle s\rangle\,J_e^2 G_e - \langle t\rangle\,J_\Delta^2 G_\Delta,
     \label{eq:resolvent3}
\\
     G_{e}^{-1} &=& z - (\langle s\rangle-1)J_e^2 G_e - \langle t\rangle \, J_\Delta^2 G_\Delta ,
     \label{eq:resolvent4}\\
       2 G_\Delta^{-1} &=& z -  (\langle t\rangle-1)J_\Delta^2 G_\Delta - \langle s\rangle J_e^2 G_e - J_\Delta ,
     \label{eq:resolvent5}
\end{eqnarray}
where we introduced the scalar quantity $G_\Delta = \boldsymbol{u}^T \mathbb{G}_\Delta   \boldsymbol{u}$. The solutions of the above equations determine the spectral
density through Eq.~(\ref{gspoq}). Equations (\ref{eq:resolvent3}-\ref{eq:resolvent5}) admit simple analytic solutions in a few particular cases. The Kesten-McKay
law for $\rho(\lambda)$ follows by setting $\langle t\rangle=0$ in the above equations \cite{Metz2011}.
For $\langle s\rangle=0$ and small values of $\langle t\rangle$, we recover the analytic expressions for the spectral density of Husimi networks \cite{Metz2011,bolle2013spectra}, while the
analytic result in \cite{Newman2019} is obtained by setting $\langle s\rangle=\langle t\rangle=1$.

Let us now present results for $\rho(\lambda)$ obtained from the numerical solutions of Eqs. (\ref{eq:resolvent3}-\ref{eq:resolvent5}). For
simplicity, we consider $J_\Delta= J_e =1$. Solving  Eq. (\ref{eq:resolvent4}) for $G_e$ and then
substituting the obtained expression into Eq. (\ref{eq:resolvent5}) yields a quartic equation for $G_\Delta$,
\begin{equation}
\big[1-\langle t\rangle-\langle s\rangle\big](\langle t\rangle-1) G_\Delta^4  + a_3 G_\Delta^3 + a_2 G_\Delta^2 + a_1 G_\Delta+ 4(\langle s\rangle-1) = 0,
\label{eq:quartic}
 \end{equation}
where the coefficients are given by
 \begin{align} \nonumber
   a_3 &=  (\langle t\rangle+z)(\langle s\rangle-2) + 2 \langle t\rangle z -2\big( \langle s\rangle - 1 \big),
\\ \nonumber
    a_2 &= (\langle s\rangle - 2)\big[\langle s\rangle + 2 (\langle t\rangle-1) \big] -\big[ (z+\langle s\rangle)-1 \big] (z-1), \\
      a_1 &= -2\Big[ z \langle s\rangle + 2 \big(1- (z + \langle s\rangle ) \big) \Big].
 \end{align}
 Hence, by solving Eq. (\ref{eq:quartic}) and inserting the result in Eqs. (\ref{eq:resolvent4}) and (\ref{eq:resolvent3}), one gets the
 spectral density from $\rho(\lambda) =  (1/\pi) \lim_{\epsilon \rightarrow 0^{+}} \textrm{Im} G(z)$.

 Figure \ref{fig:regular_net} shows the theoretical curve for $\rho(\lambda)$ obtained from the numerical solution
 of Eq.~(\ref{eq:quartic}) for $\langle s\rangle = 1$ and $\langle t\rangle=2$, along with the histogram of the eigenvalues calculated by numerically diagonalizing
 the corresponding ensemble of adjacency matrices of regular clustered graphs. The agreement between our theoretical results and numerical diagonalization data is excellent, confirming
 the exactness of Eqs. (\ref{eq:resolvent3}-\ref{eq:resolvent5}).

In Figure~\ref{fig:regular_net1}, we illustrate the shape of the spectral density derived from the solutions of Eq.~(\ref{eq:quartic}) for different
combinations of $\langle t\rangle$ and $\langle s\rangle$. We compare the Kesten-McKay law for locally tree-like
networks with degree $c=3$ and the spectral density of regular clustered networks with $\langle s\rangle=\langle t\rangle=1$. Although both network models have the same
degree $c=3$, the neighbourhood around each node of the clustered network is coupled by an additional edge.
This feature introduces pairwise correlations, which remarkably modify the eigenvalue statistics by inducing a finite gap in the spectral density.
\begin{figure}
    \centering
 \includegraphics[scale=0.55]{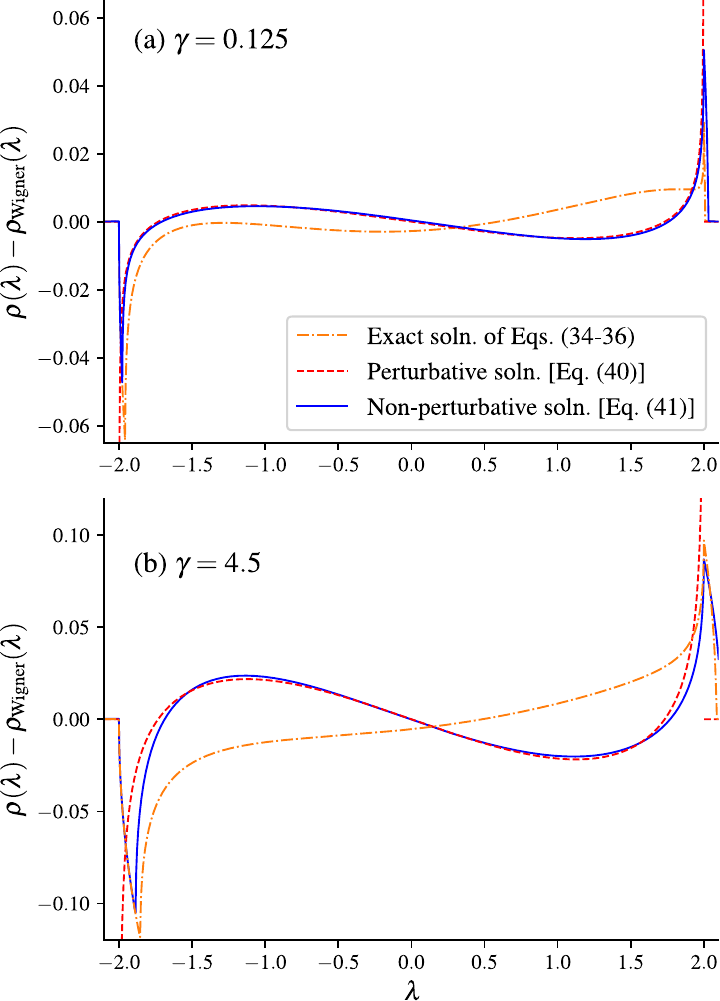}
 \caption{Deviation of the spectral density $\rho(\lambda)$ of regular clustered graphs with respect to the Wigner
   law $\rho_{\rm w}(\lambda)$ for $c=60$, $J_e=J_\Delta=1/\sqrt{c}$ and two values of the ratio $\gamma=\langle t\rangle/\langle s\rangle$. The graph compares the exact solution obtained
   from Eqs. (\ref{eq:resolvent3}-\ref{eq:resolvent5}) with the approximations given by Eqs. (\ref{eq:perturbative_solution}) and (\ref{eq:cubiclimit}).
   The values of $\lambda$ that locate the cusps identify the spectral edges obtained from the solutions of the corresponding equations.
 }
 \label{fig:deviation_from_Wigner}
\end{figure}
In the limit $c \rightarrow \infty$, $\rho(\lambda)$ converges to the Wigner
law $\rho_{\rm w}(\lambda)$ of Gaussian random matrix theory, regardless of the number $\langle t\rangle$ of triangles.
Indeed, by setting $J_e=J_\Delta=J/\sqrt{c}$
in Eqs. (\ref{eq:resolvent3}-\ref{eq:resolvent5}) and  then taking the limit $c \rightarrow \infty$, we find
\begin{equation}
    \rho_{\rm w}(\lambda) = \frac{\sqrt{\lambda_c^2-\lambda^2}}{2\pi J^2} \mathbf{1}_{(-|\lambda_c|, |\lambda_c|)}(\lambda),
    \label{eq:wigner}
\end{equation}
where $|\lambda_c|= 2 J$ is the spectral edge. The indicator function $\mathbf{1}_{(-|\lambda_c|, |\lambda_c|)}(\lambda)$ equals one if $\lambda \in (-|\lambda_c|, |\lambda_c|)$, and
zero otherwise. For large $c$, the functional form of $\rho(\lambda)$ is very close to the Wigner law $\rho_{\rm w}(\lambda)$, as can be noticed
from Figure~\ref{fig:regular_net1}.
This hints at the possibility
of devising approximate schemes for the spectral density in the large-connectivity limit.
A standard approximation technique consists in computing corrections
to $\rho_{\rm w}(\lambda)$ in a perturbative way \cite{Baron2023}.
By expanding $G(z)$ in
powers of $1/\sqrt{c}$, we find a perturbative formula for large $c$,
\begin{equation}
\rho(\lambda) \simeq \rho_{\rm w}(\lambda) + \frac{1}{\sqrt{c}} \left[ \frac{\gamma }{ \pi J^3 \left( 1+2\gamma \right)} \frac{\lambda(\lambda^2-3J^2)}{\sqrt{4J^2-\lambda^2}} \right],    
\label{eq:perturbative_solution}
\end{equation}
with $\gamma=\langle t\rangle/\langle s\rangle$. The above expression provides the leading correction to $\rho(\lambda)$.
An alternative non-perturbative approximation for the spectral density can be derived by reducing Eqs. (\ref{eq:resolvent3}-\ref{eq:resolvent5})
to a single cubic equation for the resolvent. This is achieved by setting $J_e=J_\Delta=J/\sqrt{c}$ in Eqs. (\ref{eq:resolvent3}-\ref{eq:resolvent5})
and then neglecting the lowest-order contributions of $\mathcal{O}(1/c)$, which yields the resolvent equation 
\begin{equation}
\frac{J^3}{(1+2\gamma)}\, G^3 - J\big(z+J\sqrt{c}\big)G^2  + \big(z\sqrt{c} + J\big)G - \sqrt{c} = 0
\label{eq:cubiclimit}
\end{equation}
for large $c$. The solutions of Eq.~(\ref{eq:cubiclimit}) recover the Wigner law when $\gamma=0$. Equations (\ref{eq:perturbative_solution})
and (\ref{eq:cubiclimit}) provide two distinct approximations for the spectral density of highly-connected clustered networks in the presence of triangles ($\gamma >0$).

Figure \ref{fig:deviation_from_Wigner} compares, for large mean degree $c$, the spectral density derived from the approximate Eqs. (\ref{eq:perturbative_solution}) and (\ref{eq:cubiclimit})
with the exact result that follows from the solutions of Eqs.~(\ref{eq:resolvent3}-\ref{eq:resolvent5}). The main difference between the perturbative and the non-perturbative
approaches is that the latter technique provides better approximations for the spectral edges in the presence of triangles, while the perturbative approach, in contrast, gives the same spectral
edges as those of the Wigner law. Despite that, both approximation schemes fail to yield the correct deviation from the Wigner law in the presence
of triangles. This suggests that $\rho(\lambda)$ is never given by the non-perturbative solutions of Eq. (\ref{eq:cubiclimit}) nor the perturbative
formula of Eq. (\ref{eq:perturbative_solution}), no matter how large we choose $c$.

 
\subsection{Heterogeneous networks}

\begin{figure*}[t!]
  \centering
  \includegraphics[width=2.0\columnwidth]{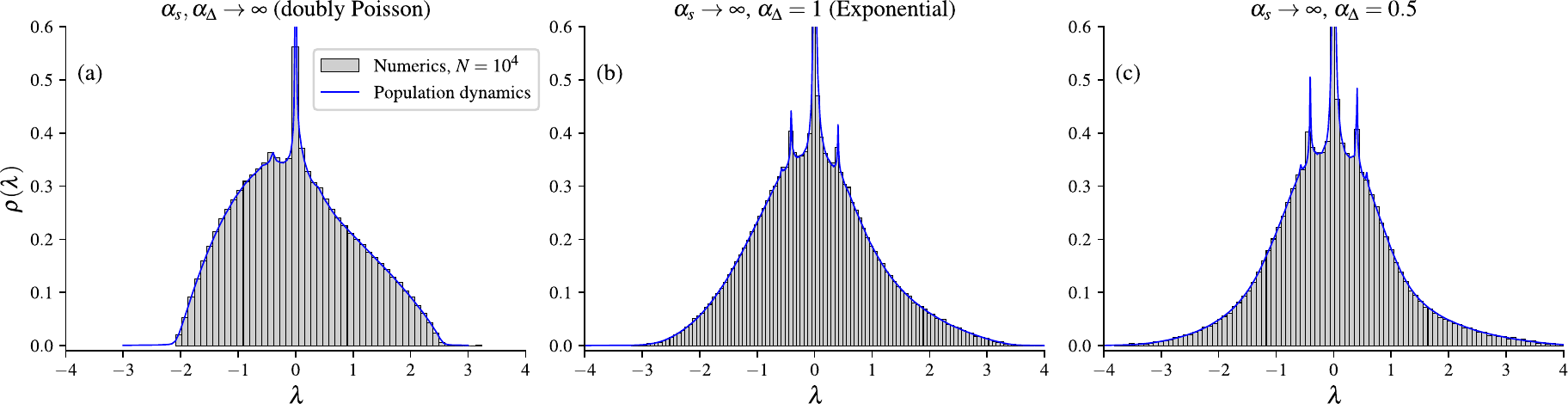}
  \caption{Spectral density of heterogeneous clustered networks with joint distribution $p_{st} = p_s^{(b)}p_t^{(b)}$ [see Eqs. (\ref{eq:neg_binom_main_text1}) and (\ref{eq:neg_binom_main_text})] for different
    combinations of the heterogeneity parameters $\alpha_e$ and $\alpha_\Delta$, whose values are indicated on each panel. All results are obtained for a fixed average number of single edges
    and triangles ($\langle s\rangle = \langle t\rangle = 2$).
    In all panels, the solid lines are calculated by solving the set of distributional
    Eqs. (\ref{green}-\ref{distributional}) using the population dynamics algorithm with $\varepsilon = 0.01$ and a total
    number of $10^5$ stochastic variables (see Appendix A for details). The histograms are obtained by numerically
    diagonalizing a single realization of the adjacency matrix of a network with $N=10^4$ nodes, randomly generated from the Newman-Miller model. }
  \label{fig4}
\end{figure*}
Our goal in this section is to investigate how the heterogeneity in the clustering shapes the spectral density of networks. To this end, we consider networks
with a joint distribution $p_{st} = p_s^{(b)} p_t^{(b)}$~(see also Appendix~\ref{sec:appendix_correlated} for results on networks with assortative and disassortative degree distributions), where the marginal distributions of the number of single-edges and triangles per node are both
given by a negative binomial distribution \cite{Metz2022,silva2022analytic}
\begin{align}
  p_{s}^{(b)}=\frac{\Gamma(\alpha_e + s)}{s! \, \Gamma(\alpha_e)} \left(\frac{\langle s\rangle}{\alpha_e}\right)^{s}\left(\frac{\alpha_e}{\alpha_e+\langle s\rangle}\right)^{\alpha_e + s}, \label{eq:neg_binom_main_text1} \\
  p_{t}^{(b)}=\frac{\Gamma(\alpha_\Delta + t)}{t! \, \Gamma(\alpha_\Delta)} \left(\frac{\langle t\rangle}{\alpha_\Delta}\right)^{t}\left(\frac{\alpha_\Delta}{\alpha_\Delta+\langle t\rangle}\right)^{\alpha_\Delta + t}, 
    \label{eq:neg_binom_main_text}
\end{align}
parametrized by the expected values $(\langle s\rangle,\langle t\rangle)$ and by the parameters $(\alpha_e,\alpha_\Delta)$. The latter are related to the variances of $p_{e}^{(b)}$ and $p_{t}^{(b)}$ as follows
\begin{equation}
    \sigma_e^2 = \langle s\rangle + \frac{\langle s\rangle^2}{\alpha_e}, \qquad  \sigma_\Delta^2 = \langle t\rangle + \frac{\langle t\rangle^2}{\alpha_\Delta}.
    \label{eq:variance_pt}
\end{equation}
We refer to clustering heterogeneity as the fluctuations in the triangle
sequence $t_1, ..., t_N$, which are controlled by the variance $\sigma_\Delta^2$.
For $\alpha_\Delta = 1$, $p_t^{(b)}$ becomes
the exponential distribution, while in the limit $\alpha_\Delta \rightarrow \infty$ it converges to the Poisson distribution. For $\alpha_\Delta \rightarrow 0$, the
variance  $\sigma_{t}^2$ diverges, and nodes with an unusually high number of triangles emerge in the network.  Therefore, by varying a single parameter, $\alpha_\Delta$, we
can continuously increase the heterogeneity in the distribution of triangles.
The same idea applies to the distribution $p_{s}^{(b)}$ of the number of single-edges.

For the standard configuration model, it has been recently shown that $\alpha_e = 1$ marks a
regime shift in the high-connectivity limit of the spectral density \cite{Metz2020,silva2022analytic}: for $\alpha_e > 1$, $\rho(\lambda)$ is a regular function, whereas it exhibits
a divergence at $\lambda=0$ when $0 < \alpha_e \leq 1$. This singular behaviour of $\rho(\lambda)$ is
ultimately caused by the multifractal structure of the eigenvectors around $\lambda=0$ \cite{Tapias_2023}. Here we shall analyze the impact of
triangle (clustering) fluctuations on $\rho(\lambda)$ under similar choices of $\alpha_\Delta$, while keeping the heterogeneity parameter $\alpha_e$ of single-edges fixed.

For heterogeneous networks, it is in general not possible to obtain closed-form solutions for the spectral density.
Thus, we numerically calculate $\rho(\lambda)$ by approximating the probability densities $W_z(g)$, $W_e(g)$ and $W_{\triangle}(\mathbb{G})$ using the population dynamics
algorithm, which is explained in Appendix~\ref{sec:appendix_population_dynamics}. In Figure~\ref{fig4} we
show results for fixed $(\langle s\rangle,\langle t\rangle)$ and three combinations of the heterogeneity parameters $(\alpha_e,\alpha_\Delta)$. In all cases, the agreement between the population dynamics
algorithm and numerical diagonalization results is excellent.

The network in the first example [Figure~\ref{fig4}(a)] has a doubly Poisson
distribution [$\alpha_e, \alpha_\Delta \rightarrow \infty$ in Eqs. (\ref{eq:neg_binom_main_text1}) and (\ref{eq:neg_binom_main_text})]. Notice that $\rho(\lambda)$ in Figure~\ref{fig4}(a) is 
asymmetric, as expected from previous works \cite{Metz2011,bolle2013spectra,Peron_2018,Newman2019}. However, if $p_s^{(b)}$ follows a
Poisson distribution ($\alpha_e \rightarrow \infty$) and we decrease the heterogeneity parameter $\alpha_\Delta$ of triangles, we see that $\rho(\lambda)$ becomes
significantly less asymmetric [Figures~\ref{fig4}(b) and \ref{fig4}(c)]. In fact, for $\alpha_\Delta = 0.5$ in panel (c), the asymmetry of $\rho(\lambda)$ is hardly
perceptible, even though the network has the same average number of triangles as in panels (a) and (b). Recall that, by decreasing $\alpha_\Delta$, we
increase the variance $\sigma_\Delta^2$ of distribution $p_t^{(b)}$ [Eq.~\eqref{eq:variance_pt}]. Therefore, the results of Figure~\ref{fig4} show
that fluctuations in the number of triangles attenuate the lack of symmetry in the spectrum of the adjacency matrix.  

\begin{figure}[t!]
	\centering
	\includegraphics[scale=0.55]{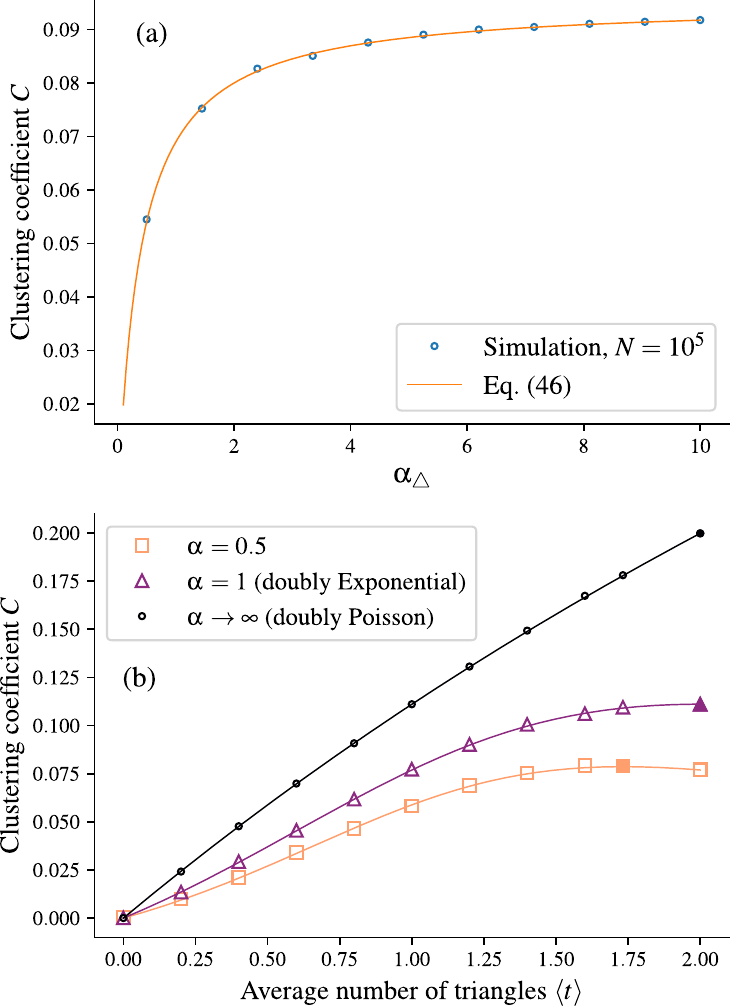}
	\caption{(a) Clustering coefficient $C$ of networks with doubly negative binomial degree distribution as a function of the
          heterogeneity parameter $\alpha_\Delta$ for fixed $\alpha_e = 2$. The average number of single-edges and triangles is $\langle s\rangle = \langle t\rangle = 2$. Each
          point is the simulation result of a single randomly generated network with $N=10^5$ nodes, while the solid line is the theoretical result
          given by Eq.~\eqref{eq:clustering_coeff_C_main}. (b) Clustering coefficient $C$ as a function of the average number of triangles, $\langle t\rangle$, for
          networks with doubly negative binomial distribution $p_{st}=p_{s}^{(b)} p_{t}^{(b)}$ with identical heterogeneity parameters $\alpha_e = \alpha_\Delta = \alpha$ and for
          fixed average degree  $c=4$. Each point is the simulation result of a single generated network with $N=10^4$ nodes, and solid lines are 
          analytic results obtained from Eq.~\eqref{eq:clustering_coeff_C_main}. Filled dots indicate the analytic
          predictions  [see Eq. (\ref{eq:ct_of_max_C_equal_alpha})] of the points
          at which the maximal value of $C$ is achieved. }
	\label{fig_clustering}
\end{figure}

To better understand the phenomenon observed in Fig.~\ref{fig4}, we calculate the clustering coefficient $C$ of networks with
a doubly negative binomial distribution $p_{st}=p_s^{(b)}p_t^{(b)}$. The clustering coefficient is defined as \cite{newman2018networks}
\begin{equation}
    C=\frac{3\times\textrm{(number of triangles)}}{\textrm{(number of connected triples)}}=\frac{3N_{\triangle}}{N_{3}},
    \label{eq:clustering_coefficient}
\end{equation}
where a triple denotes three vertices connected by two edges. 
Following the steps in Appendix~\ref{sec:appendix_clustering_coeff}, we obtain $C$ in terms
of the network parameters $\alpha_e$, $\alpha_\Delta$, $\langle s\rangle$ and $\langle t\rangle$:
\begin{equation}
    C=\frac{2\langle t \rangle}{2\langle t \rangle+\big(\langle s \rangle+2\langle t \rangle\big)^{2}+\frac{\langle s \rangle^{2}}{\alpha_{e}}+\frac{4\langle t \rangle^{2}}{\alpha_\Delta }}\,.\label{eq:clustering_coeff_C_main}
\end{equation}
Notice that for $\alpha_{e},\alpha_\Delta \rightarrow 0$, we get $C \rightarrow 0$. On the other hand, when $\alpha_{e},\alpha_\Delta \rightarrow \infty$, we recover the result
of the doubly Poisson clustered graph \cite{Newman2009}. In Figure~\ref{fig_clustering}(a), we show $C$ as a function of $\alpha_\Delta$ for networks with
fixed $\alpha_e = 2$ and $\langle s\rangle = \langle t\rangle = 2$. Note that the clustering coefficient $C$ drops as the triangle fluctuations increase ($\alpha_\Delta \rightarrow 0$). This 
occurs because the number of possible triples diverges with the variance $\sigma_\Delta^2$, while the average number of triangles remains constant across
different values of $\alpha_\Delta$. Thus, the proportion of triangles decreases as $\alpha_\Delta \rightarrow 0$, which intuitively explains the gradual loss of asymmetry in
the spectral density seen in Figure~\ref{fig4}. 

At last, we present results that illustrate how heterogeneous distributions of triangles impact
the maximal value of the clustering coefficient in the special case $\alpha_\Delta=\alpha_e=\alpha$.
By setting $\alpha_\Delta=\alpha_e=\alpha$ in Eq. (\ref{eq:clustering_coeff_C_main}), one can find the average number of triangles $\langle t\rangle=\langle t\rangle_{\max}$
that gives the maximal clustering coefficient $C_{\max}$, namely
\begin{equation}
(\langle t\rangle_{\max},C_{\max})=\begin{cases}
\left(\frac{c}{2}\sqrt{\frac{(1+\alpha)}{2}},\Sigma(c,\alpha)\right) & \textrm{for }0<\alpha<1,\\
\left(\frac{c}{2},\frac{1}{1+c (1 + \frac{1}{\alpha})}\right), & \textrm{for }\alpha \geq 1,
\end{cases}
\label{eq:ct_of_max_C_equal_alpha}
\end{equation}
where $$\Sigma(c,\alpha)\equiv\frac{\sqrt{(1+\alpha)/2}}{  (1-2c/\alpha) \sqrt{(1+\alpha)/2}+4c (1+1/\alpha)}\,.$$
For $\alpha \geq 1$, $C$ is a monotonic function of $\langle t\rangle$ and the maximal clustering coefficient is attained when $\langle t\rangle = c/2$, that is, when there are only triangles and no single-edges
in the network. Moreover, when $\alpha \rightarrow \infty$, we recover $C_{\max} = 1/(1+c)$, which is the maximum clustering coefficient for doubly Poisson
random networks. However, for $\alpha<1$, $C$ is a non-monotonic function of $\langle t\rangle$, and $C_{\max}$ is reached when the connections are not
entirely due to triangles. In other words, for networks with a heterogeneous distribution of triangles, the maximal clustered configuration is obtained
for an optimal $N_\triangle$, lower than the total number of possible triangles. This phenomenon is illustrated in Figure~\ref{fig_clustering}(b), where
we show $C$ as a function of $\langle t\rangle$ for $\alpha_\Delta=\alpha_e=\alpha$. Notice
that for doubly Poisson ($\alpha \rightarrow \infty$) and exponential ($\alpha = 1$) networks, $C_{\max}$ is obtained at the highest value of
the average number of triangles ($\langle t\rangle = 2$). For $\alpha = 0.5$, however, the maximum is found at $\langle t\rangle \simeq 1.75$.

 \section{Discussion}
\label{sec:discussion}
 
In this paper, we have used the cavity method to  derive  a closed set
of equations for the  spectral density of  random clustered networks with an arbitrary joint distribution $p_{st}$ of the
number of single-edges and triangles per node.
By numerically solving these equations using the population dynamics algorithm, one can exactly compute the spectral density for different
choices of the distribution $p_{st}$ in the limit $N \rightarrow \infty$ ($N$ is the total number of nodes).
Our theoretical findings are in excellent agreement with numerical diagonalization results for both
homogeneous (regular) and heterogeneous \emph{sparse} networks.

For networks with a doubly negative binomial distribution $p_{st}$, we have shown how
increasing fluctuations in the number of triangles reduces the
asymmetry of the spectral density. This reduction is a direct consequence of a decrease in the clustering coefficient $C$
as a function of the variance of the number of triangles per node.
Interestingly, we have also demonstrated
that $C$ reaches a maximum for a number $N_{\Delta}$ of triangles that is lower than the total number of possible triangles in the network.

For regular clustered networks, we have reduced the
problem of finding the spectral density to solving a quartic equation. In this case, the spectral density converges to the Wigner
law of random matrix theory as the mean degree increases. We have developed a perturbative and a non-perturbative approximation
for the spectral density for large mean degrees, but both approximation schemes fail to accurately predict the spectrum. This breakdown of high-connectivity
approximations occurs due to the presence of triangles, which might hint at a fundamental difficulty that arises in networks with higher-order
structures.

The exactness of our approach relies on the local tree-like property that holds
at a larger scale, similar to the cavity approach as applied to Husimi graphs \cite{Metz2011}. In principle, our method can be
extended to networks
composed of higher-order structures, as long as the neighbourhood of each node contains the same species of motifs.
Unfortunately, this is often not the case in real-world networks, where our approach is less effective. In this regard, the cavity method is more suitable
to study the spectral density of synthetic
network models, such as those variations of the configuration model that incorporate
more complex motifs (e.g.~\cite{karrer2010random,ritchie2014higher,ritchie2016beyond}), where the structure of the motifs around each node are known {\it a priori}.

From a technical perspective, our approach is equivalent to the first-order approximation
in references \cite{Newman2019,Cantwell2019}, where two neighbors of a node are correlated through a direct link, while  all higher-order
correlations coming from longer paths connecting these two neighbors are neglected. 
More specifically, the method of \cite{Newman2019,Cantwell2019} 
constructs a hierarchy of neighbourhoods around each node and then counts the number of  
closed walks that initiate and end at this node.
The first level of this hierarchical approximation assumes that the network is formed
only by single-edges and triangles, which already makes up a very good approximation
for real-world networks \cite{Cantwell2019}. In comparison to references \cite{Newman2019,Cantwell2019},  our approach is technically simpler, as
it does not involve the complicated combinatorics
of counting the closed paths of a network. In fact, while the approach in \cite{Newman2019,Cantwell2019} holds for single graph instances, our distributional equations are valid for an ensemble of networks in the limit $N \rightarrow \infty$. As such, our equations only depend on the statistical properties of the network ensemble. Even though our distributional equations could in principle be derived by taking the ensemble average of the results in \cite{Newman2019}, the connection between both methods is not straightforward, due to the intricate combinatoric structure of the equations in \cite{Newman2019}.

The resolvent equations obtained in our work pave the way to the investigation of the effect of triangle fluctuations
on the inverse participation ratio and the local density of states of sparse adjacency matrices \cite{metz2010localization}. These
quantities provide valuable information about the localization properties of eigenvectors. The resolvent equations can also be adapted to compute the spectral density
of networks with random weights, which play a central role
in the stability analysis of dynamical systems \cite{Mambuca}. Finally, it would be interesting to solve
Eqs.~(\ref{green}-\ref{distributional}) in the high-connectivity limit and probe the universality of
the Wigner law with respect to fluctuations in the number of triangles \cite{Metz2020,silva2022analytic}.
We hope that our results stimulate further work along these lines.
 
 
\begin{acknowledgments}
  Tuan Pham acknowledges support from Novo Nordisk Foundation and would like to thank Albert Alonso  for stimulating discussions. Thomas Peron acknowledges FAPESP (Grant No.~2023/07481-6) and CNPq/Brazil (Grant No. 310248/2023-0). F. L. M. acknowledges CNPq/Brazil 
  for financial support. This research was carried out using
  the computational resources of the Center for Mathematical Sciences Applied to Industry (CeMEAI) funded by FAPESP (Grant No. 2013/07375-0).
\end{acknowledgments}


\appendix

\section{Clustering coefficient of networks with joint negative binomial degree distribution}
\label{sec:appendix_clustering_coeff}

In this appendix, we calculate the clustering coefficient for networks with $p_{st} = p_s^{(b)}p_t^{(b)}$, where the marginal
distributions of single-edges and triangles are both given by  negative binomial distributions
\begin{align}
  p_{s}^{(b)}=\frac{\Gamma(\alpha_e + s)}{s! \, \Gamma(\alpha_e)} \left(\frac{\langle s\rangle}{\alpha_e}\right)^{s}\left(\frac{\alpha_e}{\alpha_e+\langle s\rangle}\right)^{\alpha_e + s}, \label{eq:neg_binom_main_appendix1} \\
  p_{t}^{(b)}=\frac{\Gamma(\alpha_\Delta + t)}{t! \, \Gamma(\alpha_\Delta)} \left(\frac{\langle t\rangle}{\alpha_\Delta}\right)^{t}\left(\frac{\alpha_\Delta}{\alpha_\Delta+\langle t\rangle}\right)^{\alpha_\Delta + t}, 
    \label{eq:neg_binom_main_appendix2}
\end{align}
where $(\langle s\rangle,\langle t\rangle)$ are the expected values  of single edges and triangles, respectively, and $(\alpha_e,\alpha_\Delta)$ are the parameters 
related to the variances of $p_s^{(b)}$ and $p_t^{(b)}$ [see Eq.~\eqref{eq:variance_pt}].

\begin{figure*}[t!]
  \centering
  \includegraphics[width=2.0\columnwidth]{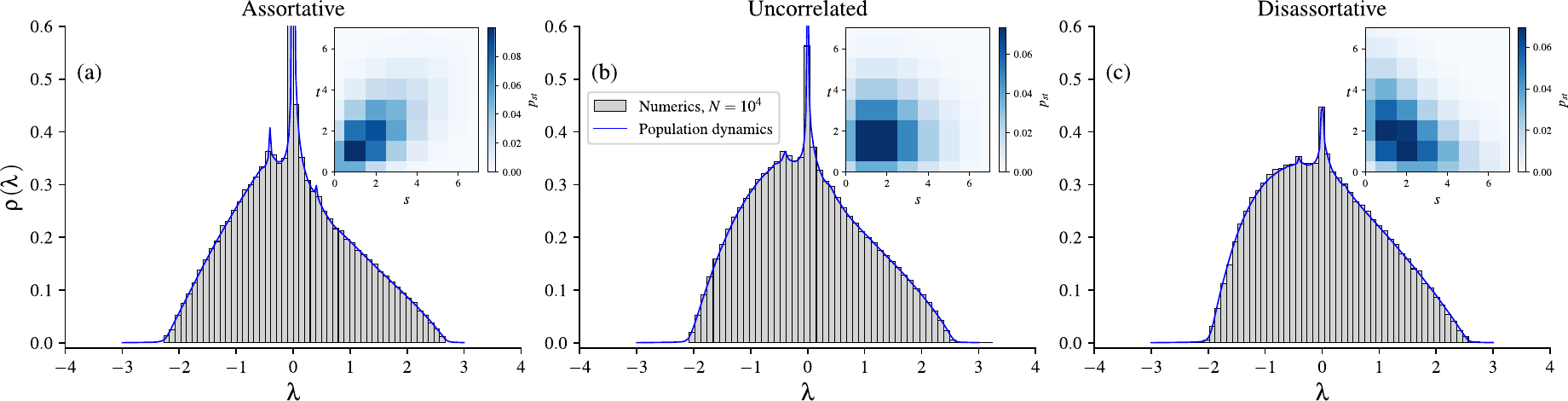}
  \caption{Spectral density of heterogeneous clustered networks with correlated degree sequences. $(a)$ Joint distribution $p_{st}$ given by Eq.~\eqref{eq:bivariate_poisson} and with parameters $\langle s \rangle=\langle t \rangle= \theta = 1$. This choice leads to a positive correlation between the number of single edges and triangles per node. In particular, the Pearson correlation coefficient for this case is $\textrm{corr}(s,t) \simeq 0.50$. $(b)$ Spectral density of networks with uncorrelated degree sequences sampled from a doubly Poisson degree distribution [same result as in Fig.~\ref{fig4}(a), which is repeated here for the sake of comparison]. $(c)$ Spectral density of networks with $p_{st}^{(\phi)}$ given by Eq.~\eqref{eq:bivariate_poisson_v2} and with parameters $\langle s \rangle = \langle t \rangle = 3$ and $\phi = 0.8$. This distribution creates networks where the number of single edges is negatively correlated with the number of triangles per node [$\textrm{corr}(s,t) \simeq -0.40$]. Insets depict the specific choice of $p_{st}$ in each case. In all panels,  the solid lines are calculated by solving the set of distributional Eqs.~(\ref{green})-(\ref{distributional}) using the population dynamics algorithm with $\varepsilon = 0.01$ and a total number of $10^5$ stochastic variables (see Appendix C for details). The histograms are obtained by numerically diagonalizing a single realization of the adjacency matrix of a network with $N=10^4$ nodes, randomly generated with the Newman-Miller model. In all panels the total average degree is $c=6$.}
  \label{fig7}
\end{figure*}

Let us start by writing the generating function of $p_{st}$:
\begin{equation}
    g_{p}(x,y)=\sum_{s,t=0}p_{st}x^{s}y^{t}.
    \label{eq:gen_func_pst}
\end{equation}
The clustering coefficient is defined as~\cite{newman2018networks}
\begin{equation}
    C=\frac{3\times\textrm{(number of triangles)}}{\textrm{(number of connected triples)}}=\frac{3N_{\triangle}}{N_{3}},
    \nonumber
\end{equation}
where a triple denotes three vertices connected by two edges. In terms of $g_p$, $N_\triangle$ and $N_3$ are written as 
\begin{equation}
    3N_{\triangle}=N\sum_{s,t=0}^{\infty}tp_{st}=N\left(\frac{\partial g_{p}}{\partial y}\right)_{x=y=1}
    \label{eq:number_of_triangles}, 
\end{equation}       
and
\begin{equation}
    N_{3}=N\sum_{k=0}^{\infty}\binom{k}{2}p_{k}=\frac{1}{2}N\left(\frac{\partial^{2}f}{\partial z^{2}}\right)_{x=y=1}, 
\label{eq:number_of_triples}
\end{equation} 
where $f(z)$ is the generating function of the total degree distribution $P_k$,
\begin{equation}
    f(z)=\sum_{k=0}^{\infty}P_{k}z^{k}=\sum_{s,t=0}^{\infty}p_{st}z^{s+2t}=g_{p}(z,z^{2}).
    \label{eq:gen_function_pk}
\end{equation}
The probability distribution in Eq.~\eqref{eq:neg_binom_main_appendix1} has a generating function given by 
\begin{equation}
f^{(b)}_e(z) = \left[\frac{\theta_e}{(1-(1-\theta_e)z)} \right]^{\alpha_e},    
\label{eq:generating_function_ps}
\end{equation}
where $\theta_e = \alpha_e/(\alpha_e + \langle s\rangle)$. The generating function $f_t^{(b)}(z)$ of $p_t^{(b)}$ is defined analogously to Eq.~\eqref{eq:generating_function_ps}, but with $\theta_{t} = \alpha_\Delta/(\alpha_\Delta + \langle t\rangle)$. Hence, $g_p(x,y)$ is written as 
\begin{equation}
g_{p}(x,y)=\left[\frac{\theta_{e}}{1-(1-\theta_{e})x}\right]^{\alpha_{e}}\left[\frac{\theta_{\triangle}}{1-(1-\theta_{\triangle})y}\right]^{\alpha_{\triangle}},
\label{eq:gen_func_negbinom}
\end{equation}
Substituting Eq.~\eqref{eq:gen_func_negbinom} into Eqs.~\eqref{eq:number_of_triangles} and~\eqref{eq:number_of_triples}, and
inserting the result in Eq.~\eqref{eq:clustering_coefficient}, we obtain the clustering coefficient 
\begin{equation}   C=\frac{2\langle t \rangle}{2\langle t \rangle+\big(\langle s \rangle+2\langle t \rangle\big)^{2}+\frac{\langle s \rangle^{2}}{\alpha_{e}}+\frac{4\langle t \rangle^{2}}{\alpha_\Delta}}.   \label{eq:clustering_coeff_C_appendix}
\end{equation}

\section{Spectral density of networks with correlated degree sequences}
\label{sec:appendix_correlated}

In this appendix, we show the spectral density of networks in which the number of single edges $s$
and triangles $t$ are correlated random variables. For positive correlations, we sample joint degree sequences $\{s_i,t_i\}_{i=1,...,N}$ from a bivariate Poisson distribution given by
\begin{equation}
    p_{st}^{(\theta)}=e^{-(\langle s\rangle+\langle t\rangle+\theta)}\frac{\langle s\rangle^{s}}{s!}\frac{\langle t\rangle^{t}}{t!}\sum_{i=0}^{\min(s,t)}\binom{s}{i}\binom{t}{i}i!\left(\frac{\theta}{st}\right)^{i}, 
    \label{eq:bivariate_poisson}
\end{equation}
where $\theta\geq 0$ is the parameter that regulates the correlation between $s$ and $t$. The total average degree of a random network sampled from $p_{st}^{(\theta)}$ is $\langle k \rangle = \langle s \rangle + 2\langle t \rangle + 3\theta$. For $\theta=0$, $s$ and $t$ become independent random variables. 

In order to generate networks with single edges negatively correlated with triangles, we consider the joint degree distribution~\cite{ghosh2021new} 
\begin{equation}
    p_{st}^{(\phi)}=K \frac{\langle s\rangle^{s}\langle t\rangle^{t}\phi^{st}}{s!t!},
    \label{eq:bivariate_poisson_v2}
\end{equation}
where $0 < \phi \leq 1 $ is the correlation parameter, and  $K$ is the normalization constant~\cite{ghosh2021new}  
\begin{equation}
    K=\left[\sum_{s=0}^{\infty}\frac{\langle s\rangle^{s}}{s!}e^{\langle t\rangle\phi^{s}}\right]^{-1}=\left[\sum_{s=0}^{\infty}\frac{\langle t\rangle^{t}}{t!}e^{\langle s\rangle\phi^{t}}\right]^{-1}.
\label{eq:constant_C_bivariate_poisson}
\end{equation}
Note that for $\phi = 1$ we recover the uncorrelated case. In Figure~\ref{fig7}, we show results for the spectral density of networks with degree distributions given by Eqs.~\eqref{eq:bivariate_poisson} and~\eqref{eq:bivariate_poisson_v2}, and fixed total average degree. As we can see, the agreement between the population dynamics algorithm (see Appendix~\ref{sec:appendix_population_dynamics}) and the numerical diagonization data is again excellent.


\onecolumngrid
\section{Algorithm for population dynamics}
\label{sec:appendix_population_dynamics}

 \begin{algorithm}[H]
\caption{Population dynamics for the calculation of $\rho(\lambda)$ of random graphs with single-edge and triangle degree sequences}
\begin{minipage}{.96\columnwidth}
\begin{algorithmic}

  \State \textbf{Input}: eigenvalue $\lambda$, regularization parameter $\varepsilon$, population size $M$, weight of single-edge
  interactions $J_e$, weight of triangle interactions $J_\Delta$, joint degree distribution $p_{st}$, number of repetitions $N_{\textrm{rep}}$.
\State \textbf{Output}: Spectral density $\rho(\lambda)$.

\State Set $p_{st}$ as the joint distribution of single edges and triangles.
\State Initialize $M$ variables $g_1,...,g_M \in \mathbb{C}$ with positive imaginary parts.
\State Initialize $M$ two-dimensional matrices $\mathbb{G}_1,...,\mathbb{G}_M \in \mathbb{C}$.
\Repeat

\State Draw two random tuples, $(s_1,t_{e1})$ and $(s_2,t_{e2})$, from $t p_{st}/\langle t \rangle$. 
\State Create two sets, $\partial_1^{(s)}$ and $\partial_2^{(s)}$, with $s_1$ and $s_2$ elements, respectively, selected uniformly at random from the indexes $1,...,M$.  
\State Create a set $\partial_1^{(t)}$ of $t_{e1}-1$ members selected uniformly at random from $1,...,M$.
\State Create a set $\partial_2^{(t)}$ of $t_{e2}-1$ members selected uniformly at random from $1,...,M$.

\State $a_{11}\leftarrow\lambda-i\varepsilon-\left( J^{2}_e\sum_{j\in\partial_{1}^{(s)}}g_{j}+J_\Delta^2\sum_{q\in\partial_{1}^{(t)}}\sum_{i,j=1}^{2}\left[\mathbb{G}_{q}\right]_{ij}\right)$  
\State $a_{22}\leftarrow\lambda-i\varepsilon-\left(J^{2}_e\sum_{j\in\partial_{2}^{(s)}}g_{j}+J_\Delta^2\sum_{q\in\partial_{2}^{(t)}}\sum_{i,j=1}^{2}\left[\mathbb{G}_{q}\right]_{ij}\right)$

\State Select an index $m_1 \in [1,M]$ uniformly at random.
\State $
\mathbb{G}_{m_1}\leftarrow\left(\begin{array}{cc} 
a_{11} & -J_\Delta\\
-J_\Delta & a_{22}
\end{array}\right) $ \Comment{Eq.~\eqref{eq:resolvent1}}
\State Select an index $m_2 \in [1,M]$ uniformly at random.

\State Draw a random tuple $(s_{e1},t_1)$ from $s p_{st}/\langle s \rangle$.

\State Create a set $\partial_s^{(3)}$ of $s_{e1}-1$ members selected uniformly at random from $1,...,M$.
\State Create a set $\partial_t^{(3)}$ of $t_{1}$ members selected uniformly at random from $1,...,M$.
\State $g_{m_{2}}\leftarrow\left[\lambda-i\varepsilon-\left(J^{2}_e\sum_{j\in\partial_{3}^{(s)}}g_{j}+J^{2}_\Delta\sum_{q\in\partial_{3}^{(t)}}\sum_{i,j=1}^{2}\left[\mathbb{G}_{q}\right]_{ij}\right)\right]^{-1}$ \Comment{Eq.~\eqref{jj1}}

\Until{step number $< N_{\textrm{rep}}$}

\For{$m=1,...,M$}
\State Draw a random tuple $(s,t)$ from $p_{st}$.
\State Create a set $\partial_s$ with $s$ elements selected uniformly at random from $1,...,M$.
\State Create a set $\partial_t$ with $t$ elements selected uniformly at random from $1,...,M$.

\State $g_{m}\leftarrow\left[\lambda-i\varepsilon-\left(J^{2}_e \sum_{j\in\partial_{s}}g_{j}+J^{2}_\Delta\sum_{q\in\partial_{t}}\sum_{i,j=1}^{2}\left[\mathbb{G}_{q}\right]_{ij}\right)\right]^{-1}$ \Comment{Eq.~\eqref{juuw}}

\EndFor

\State $\rho(\lambda)\leftarrow\frac{1}{\pi M}\textrm{Im}\sum_{j=1}^{M}g_{j} $

\Return $\rho(\lambda)$
\end{algorithmic}
\end{minipage}
\end{algorithm}

 \twocolumngrid

\bibliography{example}
 \end{document}